\documentclass[12pt]{article}
\usepackage{geometry}             
\usepackage{graphicx}
\usepackage{amssymb}
\usepackage{amsmath}
\usepackage{epstopdf}
\usepackage{comment}
\usepackage{cite}
\usepackage{abstract}

\usepackage[hyperindex=true,
          pdfstartview=FitH,
          bookmarksnumbered=true,
          bookmarksopen=true,
          citecolor=blue,
          linkcolor=blue,
          colorlinks=true,
          unicode]{hyperref}

\def\be{\begin{equation}}
\def\ee{\end{equation}}

\def\>{\rangle} 
\def\<{\langle} 
  
\usepackage{bm}

\begin{document}

\title{Holographic entanglement entropies for Schwarzschild and Reisner-Nordstr\"om black holes in asymptotically Minkowski  spacetimes}
\author{Yuan Sun and Liu Zhao\\
School of Physics, Nankai University,
Tianjin 300071, China\\
\em email:
\href{mailto: sunyuan14@mail.nankai.edu.cn}{sunyuan14@mail.nankai.edu.cn}\\
and \href{mailto: lzhao@nankai.edu.cn }{lzhao@nankai.edu.cn}
}

\date{}                             
\maketitle

\begin{abstract}
The holographic entanglement entropies (HEE) associated with four dimensional Schwarzschild
and Reisner-Nordstr\"om black holes in asymptotically Minkowski  spacetimes are investigated. Unlike the cases of asymptotically AdS
spacetimes for which the boundaries are always taken at (timelike) conformal infinities,
we take the boundaries at either large but finite radial coordinate (far boundary) 
or very close to the black hole event horizons (near horizon boundary). 
The reason for such choices is that such boundaries are similar to
the conformal infinity of AdS spacetime in that they are all timelike, so that there may be some
hope to define dual systems with ordinary time evolution on such boundaries. 
Our results indicate that, in the case of far boundaries, the leading order contribution to the 
HEEs come from the background Minkowski spacetime, however, the next to leading order 
contribution which arises from the presence of the black holes is always proportional to the 
black hole mass, which constitutes a version of the first law of the HEE for asymptotically flat
spacetimes, and the higher order contributions are always negligibly small. In the case of near 
horizon boundaries, the leading order contribution to the HEE is always proportional to the area
of the black hole event horizon, and the case of extremal RN black hole is distinguished from the 
cases of non-extremal black holes in that the minimal surface defining the HEE is completely 
immersed inside the boundary up to the second order in the perturbative expansion.
\end{abstract}

\section{Introduction}

Holography has now been widely accepted as one of the fundamental properties of 
relativistic gravitational theories, largely due to the discovery of AdS/CFT duality \cite{hep-th/9711200}.
In its preliminary form, holographic principle was originally proposed by 't Hooft \cite{gr-qc/9310026} and Susskind \cite{hep-th/9409089}
respectively, which simply states that the microscopic degrees of freedom 
of a black hole reside solely on its horizon surface. In AdS/CFT, however, the holographic
surface is promoted to the boundary at conformal infinity. Certainly this promotion has 
important significance and physical consequences. Since the boundary at conformal infinity
for asymptotically AdS spacetime is a timelike hypersurface, the dual theory on the boundary
behaves as a field theory defined on a co-dimension one spacetime. In particular, it is a theory 
sitting at the fixed point of the renormalization group, i.e. a conformal field theory. 
One can use the holographic duality to study the properties of the dual theory by 
looking at the corresponding features in the bulk or vice versa. 

Despite its great success, one should keep in mind that AdS/CFT is not the full story for
holography. The study of holographic properties of non-asymptotically 
AdS spacetimes have not been promoted to the same level of depth. One reason for this lies in 
that, for non-asymptotically AdS spacetimes, the boundary at conformal infinity is usually not a 
timelike hypersurface and hence it is difficult to define the dual theory as a normal field 
theory. To this end let us mention that for asymptotically flat spacetimes, the 
conformal infinity is always lightlike, which is very different from a timelike hypersurface.
There are other reasons, for instance, even if one can manage to define a dual theory, 
it will be usually nonlocal \cite{1010.3700} and lacks physical interests.  

Even though there are various difficulties while considering the holographic properties of 
non-asymptotically AdS spacetimes, they can never be enough to serve as a mind stopper. 
Various attempts have been practiced towards this problem  \cite{1609.06203,hep-th/9906022,
hep-th/0106113,1603.05250,1611.02702,1410.4089,1010.3700}. Among them, the holographic 
entanglement entropy (HEE) seems to be particularly suitable for tackling this problem.
Roughly speaking, the HEE is a shortcut for calculating
the entanglement entropy of a subsystem in the dual theory. When the bulk theory is 
Einstein gravity, Ryu and Takayanagi conjectured that the entanglement entropy of a subsystem
on the boundary is identical to the area (divided by $4G$) of a minimal surface which takes the 
entanglement surface as its own boundary \cite{hep-th/0603001,hep-th/0605073}. {{This 
conjecture has been proven in \cite{1304.4926}, and also generalized to time-dependent background 
\cite{0705.0016} and higher derivative gravity \cite{1310.5713}. }}

Since RT formula involves the minimal surface, it is important to analyze such minimal surfaces 
in various asymptotically AdS spacetime with black hole in the bulk or time-dependent background, 
for instance in \cite{1306.4004,1312.6887}. Extremal surfaces are also analyzed in de 
Sitter spacetime \cite{1501.03019,1504.07430}. In this paper, we shall take 
Ryu-Takayanagi conjecture as the definition of HEE 
in asymptotically flat spacetimes \cite{1010.3700}. More concretely, we shall study the 
HEE in four dimensional Schwarzschild and Reisner-Nordstr\"om (RN) spacetimes 
and analyze its dependence on various impacting parameters. To avoid getting only divergent
results, we do not put the boundary at conformal infinity. Instead, we take the boundary 
at some finite radial position which is either very far from the black hole horizon or
is very close to the horizon. Such boundary hypersurfaces are all timelike, and therefore
there leaves room for defining normal dual field theories on such hypersurfaces. 
However, remember that the boundary hypersurfaces we take are not sitting at the 
renormalization group fixed point, so even there exist normal dual field theories, these 
will not be conformal. Such theories may just be some off critical theories. 

The rest of this paper is organized as follows. In Section 2, we study the 
HEE in Schwarzschild and RN spacetimes with far boundaries. It will be shown that the behavior
of the HEE for both spacetimes are quite similar. The difference begins to show up
only at the second order in the perturbative expansion, which is negligibly small.
In Section 3, we study the same problem but with the far boundaries replaced by near 
horizon ones. The HEE behaves differently for the case of extremal RN spacetime in contrast to 
the cases of Schwarzschild and non-extremal RN spacetimes.  The HEEs in all cases will be
calculated up to the second order in the perturbative expansion and will be cross checked 
by numerical procedures. In the last section, Section 4, we summarize some of the interesting
results obtained in this paper.  

\section{HEE in Schwarzschild and RN spacetimes -- far boundary}

In this section, we shall study the HEE in Schwarzschild and RN spacetimes with the dual 
theory defined on a boundary located at large but finite radial coordinate $r=r_\infty$. 
The bulk theory is assumed to be Einstein gravity and the spacetime metric takes the well 
known form
\begin{align}
\mathrm{d}s^2&=-A(r)\mathrm{d}t^2+B(r)\mathrm{d}r^2
+r^2(\mathrm{d}\theta^2+\sin^2\theta \mathrm{d}\varphi^2),\nonumber\\
A(r)&=B(r)^{-1}=1-\frac{2M}{r}+\frac{Q^2}{r^2}. \label{spacetime}
\end{align}
When both of the parameters $M$ and $Q$ 
equal to zero, the metric becomes that of the Minkowski spacetime. When $M\neq 0$ but $Q=0$, 
the metric corresponds to Schwarzschild spacetime. And when both $Q$ and $M$ are nonzero, 
the metric corresponds to RN spacetime. With either $M\neq 0$ or both $M$ and $Q$ are nonzero,
we can think of the contributions of the terms involving $M$ and/or $Q$ as perturbations 
at large enough radial distances.

In this section, we take the boundary hypersurface at $r=r_\infty$ with 
$r_\infty$ very large but still finite. The entanglement surface will be taken to be a circle 
characterized by $\theta=\theta_0$, and the minimal surface 
which takes the entanglement surface as its boundary can be described by the radial
coordinate $r$ as a function of $\theta$, i.e. $r=r(\theta)$, which is to be determined 
via the minimization of the area function
\be
\mathcal{A}={2\pi}\int_0^{\theta_0}\mathrm{d}\theta \,r\sin\theta
\sqrt{B(r)\left(\frac{\mathrm{d}r}{\mathrm{d}\theta}\right)^{2}+r^2}, \label{area}
\ee
with the boundary condition $r(\theta_0)=r_\infty$.
Once the above variational problem is solved, the HEE will be
given by the Ryu-Takayanagi formula
\be 
S=\frac{\mathcal{A}}{4G}. \label{RT}
\ee

First let us consider the HEE in the Minkowski background. In this case, we have $B=1$ 
and the minimal surface is simply a flat disk with equation 
$z_0=r_\infty\cos\theta_0=r\cos\theta$. This surface has area $\pi r_\infty^2\sin^2\theta_0$, 
hence the HEE is \cite{1010.3700}
\be \label{leadingEE}
S_0=\frac{\pi r_\infty^2\sin^2\theta_0}{4G}=\frac{\pi z_0^2\tan^2\theta_0}{4G}.
\ee

Now let us return to the case with generic $M$ and $Q$. For simplicity we denote $x=\cos\theta$.
Then the area formula \eqref{area} can be rewritten as
\be \label{Lag1}
\mathcal{A}=\int_{x_0}^1 \mathrm{d}x \, L 
= {2\pi}\int_{x_0}^1 \mathrm{d}x \, r\sqrt{B(r)(1-x^2)r'^2+r^2},
\ee
where $x_0=\cos\theta_0$ and the prime denotes derivative with respect to $x$. Variation of 
\eqref{Lag1} with respect to $r(x)$ yields the equation
\be  \label{EOM}
(x^2-1)\left[2Br^2r''-2xB^2r'^3+\left(r\frac{\mathrm{d}B}{\mathrm{d}r}-6B\right)rr'^2\right]
+4xBr^2r'+4r^3=0.
\ee
When $B[r(x)]=1$, one can check that $r= z_0/x$ is a solution. 
The equation \eqref{EOM} is highly nonlinear. In order to get a nontrivial solution 
with nonzero $M$ and $Q$, let us expand $B[r(x)]$ and $r(x)$ into formal series,
\be 
B[r(x)]=1-\sum_{n\geq 1} f_n(x)\epsilon^n, \label{bep}
\ee
\be 
r(x)=\frac{z_0}{x}+\sum_{n\geq 1}r_n(x)\epsilon^n, \label{rep}
\ee
where $\epsilon = \frac{M}{r_\infty}$ is a small dimensionless constant and all terms in 
positive powers of $\epsilon$ in eqs. \eqref{bep} and \eqref{rep} represent modifications
arising from the presence of black hole. The absence of $O(\epsilon^0)$ term in \eqref{rep}
can be understand from the background solution $B[r(x)]=1, r(x)=z_0/x$.
Using the full form of $B[r(x)]$ presented in \eqref{spacetime}, it is not difficult to 
get
\begin{align}
f_1(x)=-\frac{2 r_\infty  x}{z_0},\quad
f_2(x)=\frac{r_\infty  x^2 \left[\left(\xi^2-4\right)
   r_\infty +2 r_1(x)\right]}{z_0^2},\quad \cdots
\end{align}
where $\xi=Q/M$ is the charge to mass ratio. For Schwarzschild spacetime, $\xi=0$. For RN 
spacetime, $\xi\neq0$, wherein $0<|\xi|<1$ 
represents a non-extremal black hole spacetime, $|\xi|=1$ represents an extremal spacetime, and 
$|\xi|>1$ corresponds to a spacetime with a naked singularity.

In the following, we shall use the above expansion to get approximate solutions to the
equation \eqref{EOM} up to the second order in $\epsilon$ and evaluate the corresponding
modifications to the HEE. Similar procedures have been applied in \cite{1512.02816} 
wherein the metric perturbation of HEE in global AdS coordinate is studied, and also in 
\cite{1610.01542}, where the time dependence of holographic entanglement complexity is
analyzed. 

\subsection{First order}

To the first order in $\epsilon$, eq. \eqref{EOM} becomes
\begin{align} \label{eq1}
r''_1+\frac{5x^2-3}{x^3-x}r'_1+\frac{3x^2-1}{x^4-x^2}r_1
=\frac{r_\infty(3x^2+1)}{x^4-x^2}.
\end{align}
With $f_1$ provided as a known function, eq. \eqref{eq1} is a second order linear inhomogeneous
differential equation in $r_1(x)$. It is known that for generic second order linear 
inhomogeneous differential equations of the form
\be
r''_1+P(x)r'_1+Q(x)r_1=G(x),
\ee
the general solution can be written as
\be 
r_1(x)=c_1u_1(x)+c_2u_2(x)+\int^xdy\frac{(u_1(y)u_2(x)-u_1(x)u_2(y))G(y)}{W(y)},
\label{gensol}
\ee
where $u_1$ and $u_2$ are solutions for the homogeneous equation
\be
r''_1+P(x)r'_1+Q(x)r_1=0,
\ee
and $W=u_1u'_2-u'_1u_2$ is the Wronskian of the solutions $u_1$ and $u_2$.

For eq.(\ref{eq1}), we can easily get 
\be \label{so1}
u_1(x)=\frac{1}{x},~~u_2=\frac{2\ln x-\ln(1-x^2)}{2x},
\ee
and hence
\be
W(x)=\frac{1}{x^3-x^5}.
\ee
Inserting $W(x)$ and 
\be
G(x)=\frac{r_\infty(3x^2+1)}{x^4-x^2}
\ee
into \eqref{gensol}, we get
\begin{align} 
r_1(x)=\frac{c_1}{x}+c_2\frac{2\ln x-\ln(1-x^2)}{2x}+
r_\infty\left(1+\frac{\ln(1-x)}{x}-\frac{\ln(1+x)}{x}\right).
\end{align}
Since $x=\cos \theta\in[\cos\theta_0,1]$, the terms containing $\ln(1-x)$ must be cancelled to 
avoid divergence, which is used to determine $c_2=2r_\infty$. $c_1$ can be fixed by the boundary 
condition $r_1(x_0)=0$, which results in
\be
c_1=r_\infty\left[-x_0-2\ln x_0+2\ln(1+x_0)\right].
\ee
Finally the first order contribution to $r(x)$ is
\be 
r_1(x)=\frac{r_\infty}{x}\left[(x-x_0)+2\ln\left(\frac{x}{x_0}\right)
+2\ln\left(\frac{x_0+1}{x+1}\right)\right].
\ee
Notice that $r_1$ given above diverges when $x=0$ (i.e. $\theta=\pi/2$), therefore it is 
necessary to restrict $\theta_0<\pi/2$, so that $x=0$ can never be approached. This is 
a limitation of the perturbative expansion in the case of far boundaries. The same limitation 
also appears in the case of AdS background \cite{1512.02816}.  

In order to evaluate the first order modification to the HEE, let us expand the 
integrand $L$ in (\ref{Lag1}) to the first order in $\epsilon$,
\be \label{sol1_fi}
L=L_0+L_1\epsilon +\cdots,
\ee
so that \eqref{Lag1} becomes
\be
\mathcal{A}=\mathcal{A}_0+\mathcal{A}_1 +\cdots,
\ee
with
\begin{align}
\mathcal{A}_0&=\int_{x_0}^1 \mathrm{d}x \,L_0= \int_{x_0}^1 \mathrm{d}x \,\frac{2\pi z_0^2}{x^3}
=\pi z_0^2\left(\frac{1}{x_0^2}-1\right),\\
\mathcal{A}_1&=\epsilon \int_{x_0}^1 \mathrm{d}x \,L_1 
= \frac{M}{r_\infty}\int_{x_0}^1 \mathrm{d}x \,\frac{4\pi r_\infty z_0}{x^3}
\left[2x-x_0-1+2\ln\left(\frac{x}{x_0}\right)
+2\ln\left(\frac{x_0+1}{x+1}\right)\right]\nonumber\\
&=2\pi M(1-x_0)^2 r_\infty.
\end{align}
It is not surprising that replacing $\mathcal{A}$ with $\mathcal{A}_0$ in \eqref{RT}
would yield the HEE \eqref{leadingEE} at the background level. Similarly, replacing 
$\mathcal{A}$ with $\mathcal{A}_1$ in \eqref{RT} yields the first order modification to the 
HEE,
\be
S_1 = \frac{\pi M(1-x_0)^2}{2G} r_\infty.
\ee
Since we assume $r_\infty$ to be very large and $S_0\propto r_\infty^2$, $S_1 \propto r_\infty$,
we can safely regard $S_1$ to be a ``small'' modification to $S_0$. 
An important point to observe is that 
\be 
\delta S\equiv S-S_0 \simeq S_1 = \frac{\pi r_\infty(1-x_0)^2}{2G} M,
\ee
i.e. the small modification to the HEE is proportional to the black hole mass. This behavior is 
similar to the case of HEE associated with black holes in AdS backgrounds \cite{1212.1164}. In the latter case,
the result $\delta S\propto M$ is known as the first law of HEE. Another point to observe lies in 
that $S_1$ is independent of the charge $Q$, so that it can not distinguish the cases of
Schwarzschild and RN black holes. This is also not surprising because we are now considering
the far boundaries, and the charge term in  \eqref{spacetime} plays as the second order 
modification in the far end. To observe the contribution of the charge parameter, we have to 
move on to the second order.

\subsection{Second order}

At the second order, eq.\eqref{EOM} takes the form  
\begin{align} \label{eq2}
r''_2+\frac{5x^2-3}{x^3-x}r'_2+\frac{3x^2-1}{x^4-x^2}r_2=G_2(x),
\end{align}
where the homogeneous part takes the same form as that of the first order equation (\ref{eq1}), 
thus the solutions of the homogeneous equation remain unchanged. The inhomogeneous 
term $G_2(x)$ takes the value
\be 
G_2(x)=\frac{2r_\infty^2\left[(\xi^2-1)x^3-3x+4\right]}{z_0 x^2(1-x^2)}.
\ee 
The solution of eq.\eqref{eq2} can be written as 
\be 
r_2(x)=c_3 u_1(x)+c_4 u_2(x)+u(x),
\ee
where $u_1,u_2$ are given in \eqref{so1}, and 
\begin{align*} 
u(x)&\equiv\int^xdy\frac{(u_1(y)u_2(x)-u_1(x)u_2(y))G_2(y)}{W(y)}\\
&=-\frac{r_\infty^2 \left(\left(\xi^2-1\right)
   x^2+\left(\xi^2+9\right) \ln
   (1-x)+\left(\xi^2-23\right) \ln
   (x+1)\right)}{4 x z_0}.
\end{align*}
Similar to the case discussed in the previous subsection, the integration constant $c_4$ can be 
determined by the cancellation of the term involving $\ln(1-x)$, and the boundary condition 
$r_2(x_0)=0$ fixes $c_3$,
\begin{align}
c_3&=\frac{r_\infty^2}{4z_0}\left[(\xi^2-1)x_0^2+2(\xi^2+9)\ln x_0-32\ln(1+x)\right],\\
c_4&=-\frac{r_\infty^2(\xi^2+9)}{2z_0}.
\end{align}
Thus we get the final expression for $r_2(x)$,
\be
r_2(x)=\frac{r_\infty^2}{4z_0 x}\left[(1-\xi^2)(x^2-x_0^2)
+2(\xi^2+9)\ln\left(\frac{x_0}{x}\right)
+32\ln\left(\frac{1+x}{1+x_0}\right)\right].
\ee

To obtain the second order modification to the HEE, we need to expand the area $\mathcal{A}$ and 
the integrand $L$ in \eqref{Lag1} up to the second order,
\begin{align}
L&=L_0+L_1\epsilon+L_2 \epsilon^2+\cdots,\\
\mathcal{A}&=\mathcal{A}_0+\mathcal{A}_1+\mathcal{A}_2+\cdots.
\end{align}
Omitting the very complicated expression for $L_2$, we present directly the final result for
$S_2\equiv \frac{\mathcal{A}_2}{4G}=\frac{\epsilon^2}{4G} \int_{x_0}^1 \mathrm{d}x \,L_2$:
\begin{align}\label{s2sch}
S_2 
&=  \frac{\pi}{8G}\left\{
\left[(1-x_0)(x_0-7)+2\ln x_0+16\ln\left(\frac{2}{1+x_0}\right)\right]M^2
+\left[1-x_0^2+2\ln x_0\right] Q^2\right\}.
\end{align}
It can be seen explicitly that the charge begins to contribute at this order, but
since $S_2/S_1\sim O(\epsilon)$, such contribution is always negligibly small. 
As a consequence, the first law for the HEE
\be
\delta S = S-S_0\propto M
\ee
holds very well even if we calculate $S$ up to the second order.

\section{HEE in Schwarzschild and RN spacetimes -- near horizon}

The study of HEE with entangling surface located on the far boundary indicates that 
there is no significant difference between the cases of Schwarzschild and RN spacetimes.
This is of course just an illusion stemming from the particular choice of boundary at 
large $r_\infty$. In this section, we shall study the opposite choice, i.e. near horizon 
boundaries. For such boundaries, the differences between the HEEs for 
Schwarzschild, non-extremal RN and extremal RN spacetimes will become more transparent. By studying near horizon geometry of AdS black hole, Carlip found the black hole entropy using Cardy's formula \cite{Carlip:1998wz}. This also motivates us to put the boundary near horizon in the case of asymptotic flat black holes. 

\subsection{Schwarzschild}

First we consider the case of Schwarzschild spacetime by setting $Q=0$ in \eqref{spacetime}.
In the near horizon limit, it is convenient to introduce a new radial coordinate  
\be
\rho=\sqrt{r-r_+},\quad  (r_+=2M)
\ee
after which the metric on the constant $t$ hypersurface becomes
\be 
\mathrm{d}s^2=4g(\rho)\mathrm{d}\rho^2
+g(\rho)^2(\mathrm{d}\theta^2+\sin^2\theta \mathrm{d}\varphi^2),
\ee
where $g(\rho)=r_++\rho^2$. In this new coordinate, the black hole event horizon is located at
$\rho=0$. 

Now let us take a near horizon boundary at $\rho=\rho_0$ with 
$\epsilon\equiv \rho_0/\sqrt{r_+}\ll 1$. Once again, the one
dimensional circle characterized by $\theta=\theta_0$ is taken as the entangling surface.
The two dimensional minimal surface which takes the above circle as its boundary 
possesses an induced metric 
\be 
\mathrm{d}\hat{s}^2=\left[4g(\rho)\left(
\frac{\mathrm{d}\rho}{\mathrm{d}\theta}\right)^2+g(\rho)^2\right]\mathrm{d}\theta^2
+[g(\rho)\sin\theta]^2\mathrm{d}\varphi^2,
\ee 
where $\rho=\rho(\theta)$ is to be determined by minimizing 
the surface area
\be \label{area}
\mathcal{A}=\int_{x_0}^1 \mathrm{d}x \mathrm{d}\varphi \, L
=2\pi \int_{x_0}^1 \mathrm{d}x g(\rho)\left[4g(\rho)(1-x^2)
\rho^{\prime 2}+g(\rho)^2\right]^{1/2},
\ee
Notice that we have made a coordinate change $\theta\to x=\cos\theta$ in eq.\eqref{area}
just like in the last subsection, and the prime denotes derivative with respect to 
$x$.

The variation of eq.(\ref{area}) yields
\be \label{EOMsch}
2g(x^2-1)\rho''+8x(1-x^2)\rho^{\prime 3}
+5(1-x^2)\frac{\mathrm{d}g}{\mathrm{d}\rho}\rho^{\prime 2}
+4g x\rho'+g\frac{\mathrm{d}g}{\mathrm{d}\rho}=0.
\ee
Direct analytical solution to \eqref{EOMsch} seems to be very difficult to find. Therefore,
we would like to find the minimal surface perturbatively following a similar procedure
as we have done in the case of far boundaries. 

By observing eq.(\ref{area}), one sees that any surface anchored on the circle at 
$\rho=\rho_0, \theta=\theta_0$ and extended into the bulk with $\rho>\rho_0$  will have an area 
greater than that of the surface lies along $\rho=\rho_0$ with the same boundary. 
In other words, if there exists a minimal surface anchored on the above circle, it must lie in 
the region $\rho<\rho_0$. Thus the expansion of $\rho(x)$ must start from the first order in 
$\epsilon$,
\be \label{exprho}
\rho(x)=\rho_1(x)\epsilon+\rho_2(x)\epsilon^2+\cdots,
\ee
and the boundary condition $\rho(x_0)=\rho_0$ becomes
\be
\rho_1(x_0) =\sqrt{r_+},\quad \rho_2(x_0) =0.  \label{bdry}
\ee

Substituting eq.(\ref{exprho}) into \eqref{EOMsch} and expanding to the first order in 
$\epsilon$, one gets   
\be 
(x^2-1)\rho''_1+2x\rho'_1+\rho_1=0. \label{Legdr}
\ee
This is a special case of the Legendre equation. The solution which is regular at $x=1$ and 
satisfies the boundary condition \eqref{bdry} reads
\be 
\rho_1(x)=c_1 P_\nu(x),~~\nu=-\mathrm{e}^{-\mathrm{i}\pi/3},
~~c_1=\frac{\sqrt{r_+}}{P_\nu(x_0)},
\ee
where $P_\nu(x)$ is the Legendre polynomial of the first kind.
Expanding to the second order in $\epsilon$, the equation for $\rho_2$ follows, 
\be 
(x^2-1)\rho''_2+2x\rho'_2+\rho_2=0.
\ee
This equation takes the same form as \eqref{Legdr}. Thus the solution can be written as
\be 
\rho_2(x)=c_2P_\nu(x),~~\nu=-\mathrm{e}^{-\mathrm{i}\pi/3},
\ee
and the boundary condition $\rho_2(x_0)=0$ requires $c_2=0$, which sets $\rho_2(x)=0$.

Inserting eq.(\ref{exprho}) into the integrand $L$ in \eqref{area}, we get 
\be 
L=2\pi r_+^2 +4\pi r_+[(1-x^2)\rho_1^{\prime 2}+\rho_1^2]\epsilon^2+\cdots.
\ee
Substituting back into \eqref{area} and expand $\mathcal{A}$ accordingly,
\be
\mathcal{A}=\mathcal{A}_0+\mathcal{A}_1+\mathcal{A}_2+\cdots,
\ee
we will have
\begin{align}
\mathcal{A}_0 = 2\pi r_+^2(1-x_0),\quad
\mathcal{A}_1=0,
\end{align}
and
\be
\mathcal{A}_2=\frac{4\pi r_+ \rho_0^2}{(P_\nu(x_0))^2}
\int_{x_0}^1 \mathrm{d}x\,\left[(1-x^2)
\left(\frac{\mathrm{d}P_\nu(x)}{\mathrm{d}x}\right)^2+(P_\nu(x))^2\right].
\ee

\begin{figure}[htbp!]
\begin{center}
\includegraphics[width=.5\textwidth]{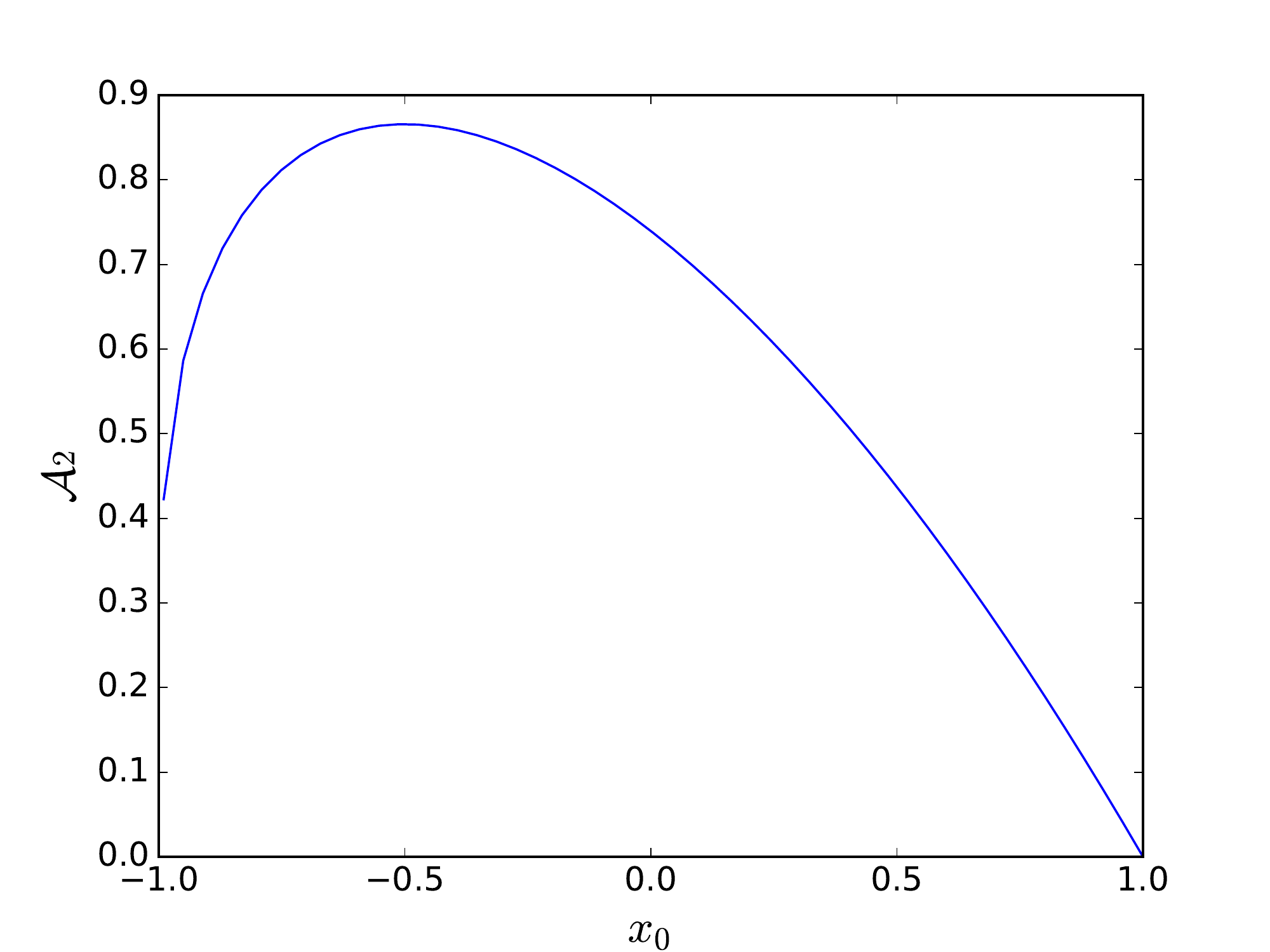}
\end{center}
\caption{ $\mathcal{A}_2$ versus $x_0$, in units of $4\pi r_+ \rho_0^2$} \label{fig1}
\end{figure}

The last integration cannot be worked out analytically, however it is easy to get the 
numerical result. It turns out that for all $-1<x_0<1$, $\mathcal{A}_2$ is finite and 
is less than 1 in units of $4\pi r_+\rho_0^2$. 
Fig.\ref{fig1} gives the plot of $\mathcal{A}_2/(4\pi r_+\rho_0^2)$ 
versus $x_0$. Since $4\pi r_+\rho_0^2$ is roughly $\epsilon^2 \mathcal{A}_0$, we can see 
that the HEE associated with Schwarzschild spacetime with a near horizon boundary is almost 
precisely given by the zeroth order contribution. In particular, the HEE equals the 
Beckenstein-Hawking entropy for the black hole when $x_0=-1$, i.e. $\theta_0=\pi$. Let us 
remark that there is no problem of divergence in the case of near horizon boundaries.

\subsection{Non-extremal RN}

Let us now turn our attention to the case of non-extremal RN black hole spacetime with a near 
horizon boundary.

For non-extremal RN black hole spacetime, the functions $A(r)$ and $B(r)$ in \eqref{spacetime}
can be written in a factorized form,
\be
A(r)=B(r)^{-1}=\frac{(r-r_-)(r-r_+)}{r^2},\quad r_\pm=M\pm\sqrt{M^2-Q^2}. \label{ABRN}
\ee
In the near horizon limit, making the coordinate transformation $r\to \rho$ with $r=r_+ +\rho^2$, 
the metric on a constant $t$-slice becomes
\be 
\mathrm{d}s^2=f(\rho)\mathrm{d}\rho^2+g(\rho)^2(\mathrm{d}\theta^2
+\sin^2\theta \mathrm{d}\varphi^2),  \label{dsrnsp}
\ee
where 
\be \label{gf}
g(\rho)=r_++\rho^2,\quad f(\rho)=\frac{4(\rho^2+r_+)^2}{\rho^2+r_+-r_-}.
\ee
The induced metric on a co-dimension 2 minimal surface is 
\be 
\mathrm{d}\hat{s}^2=\left[f(\rho)(1-x^2)\rho^{\prime 2}+g(\rho)^2\right]
\mathrm{d}\theta^2+g(\rho)^2\sin^2\theta \,\mathrm{d}\varphi^2, 
\label{nern-A}
\ee
where again $x=\cos\theta$ and primes denote derivatives with respect to $x$. 
The area functional is then 
\be 
\mathcal{A}=\int \mathrm{d}x\mathrm{d}\varphi\, L
=2\pi \int \mathrm{d}x\, g(\rho)
[f(\rho)(1-x^2)\rho^{\prime2}+g(\rho)^2]^{1/2}, \label{mins}
\ee 
and the minimization condition yields the following differential equation:
\be \label{eomrn}
(1-x^2)\left[-2fg^2\rho''+2f^2x\rho^{\prime3}
-\left(\frac{\mathrm{d}f}{\mathrm{d}\rho}g^2-6fg\frac{\mathrm{d}g}{\mathrm{d}\rho}\right)
\rho^{\prime2}\right]
 +4fg^2x\rho' +4g^3\frac{\mathrm{d}g}{\mathrm{d}\rho}=0.
\ee
Let us again take the small parameter $\epsilon\equiv \rho_0/\sqrt{r_+}\ll 1$ and make the 
expansion
\be
\rho(x)=\rho_1\epsilon+\rho_2\epsilon^2+\cdots.  \label{exp-rho}
\ee
In the first order, eq.\eqref{eomrn} can be reduced into
\be
(x^2-1)\rho_1''+2x\rho'_1+\frac{r_+-r_-}{r_+}\rho_1=0.
\ee
Once again, we get a specific Legendre equation and the general solution reads 
\be
\rho_1(x)=c_1 P_\nu(x)+c_2 Q_\nu(x),\quad \nu=\sqrt{\frac{r_-}{r_+}-\frac{3}{4}}-\frac{1}{2}.
\ee
Here $Q_\nu(x)$ is the Legendre polynomial of the second kind, and
$0< r_-/r_+< 1$. The solution regular at $x=1$ for generic choice of $r_\pm$ is 
\be 
\rho_1(x)=c_1P_\nu(x),~~c_1=\frac{\sqrt{r_+}}{P_\nu(x_0)},
\ee
where $c_1$ is determined by the boundary condition $\rho_1(x_0)\epsilon=\rho_0.$

The equation at the next order takes the same form as that for $\rho_1$, i.e.
\be
(x^2-1)\rho_2''+2x\rho'_2+\frac{r_+-r_-}{r_+}\rho_2=0.
\ee
Then the the boundary condition $\rho_2(x_0)=0$ forces $\rho_2(x)=0$, just as in the 
Schwarzschild case.

Now let us expand the integrand $L$ in \eqref{nern-A} up to the second order in $\epsilon$,
\begin{align}
L&=2\pi r_+^2+\frac{4\pi r_+^2}{r_+-r_-}\left((1-x^2)\rho_1^{\prime 2}
+\frac{r_+-r_-}{r_+}\rho_1^2
\right)\epsilon^2+
\cdots.
\end{align}
Then the area of the minimal surface is
\begin{align}
\mathcal{A}&=2\pi(1-x_0)r^2_++\frac{4\pi r_+^2}{r_+-r_-}\int_{x_0}^1\mathrm{d}x 
\left((1-x^2)\rho_1^{\prime 2}
+\frac{r_+-r_-}{r_+}\rho_1^2\right)\epsilon^2+ \cdots\nonumber\\
&=2\pi(1-x_0)r^2_++\frac{4\pi r_+ \rho_0^2}{r_+-r_-}\int_{x_0}^1\mathrm{d}x 
\left((1-x^2)\rho_1^{\prime 2}
+\frac{r_+-r_-}{r_+}\rho_1^2\right)+ \cdots.
\label{arne}
\end{align}
When $r_-=0$, this reduces to the corresponding result for the Schwarzschild case. 
Note, however, that the second order term contains $r_+-r_-$ in the denominator,
therefore this result is not applicable to the case of extremal RN black hole 
(i.e. the $r_+=r_-$ case).

\subsection{Extremal RN}

For extremal RN black hole spacetime, \eqref{ABRN} becomes
\be
A(r)=B(r)^{-1}=\frac{(r-r_+)^2}{r^2},\quad r_+=M. \label{ABRN2}
\ee
Making a coordinate change $r\to\rho$ with $r=r_++\rho^2$, the metric on the 
constant $t$-slice of the spacetime reads
\be 
\mathrm{d}s^2=f(\rho)\mathrm{d}\rho^2+g(\rho)^2(\mathrm{d}\theta^2
+\sin^2\theta \,\mathrm{d}\varphi^2),
\ee
with 
\be \label{gf}
g(\rho)=r_++\rho^2,\quad f(\rho)=\frac{4(\rho^2+r_+)^2}{\rho^2}.
\ee

Formally, this metric is identical to \eqref{dsrnsp}, the only difference lies in that
we have now a different function $f(\rho)$ in the first term on the right hand side. 
Repeating the process as in \eqref{nern-A}-\eqref{eomrn} and carrying out 
the expansion \eqref{exp-rho}, we can get a very complicated equation 
with a denominator of order $O(\epsilon)$. Discarding this denominator and looking only
at the numerator, we can get, at the leading order $O(\epsilon^0)$, the following equation
for $\rho_1(x)$,
\be \label{rho12}
2 x \rho_1^2 \rho_1 '+\left(1-x^2\right) \left(- \rho_1^2 \rho_1 ''
+4 x \rho_1^{\prime 3}+ \rho_1 \rho_1^{\prime 2}\right) =0.
\ee
This is a nonlinear homogeneous differential equation in which every term is cubic in the 
function $\rho_1(x)$. This equation can have many analytical solutions, but the only solution
which satisfies the boundary condition $\rho_1(x_0)\epsilon =\rho_0$ and is regular at $x=1$
is a constant solution
\be
\rho_1(x)= \rho_0/\epsilon=\sqrt{r_+}.
\ee
We can evaluate the numerator of \eqref{eomrn} to the next 
few orders in $\epsilon$ and find that the only solution for $\rho_2(x)$ which obeys the boundary 
condition $\rho_2(x_0)\epsilon^2=0$ is simply $\rho_2(x)=0$. Thus we have $\rho(x)=
\rho_1(x)\epsilon=\rho_0$, i.e. the minimal surface is completely immersed inside the 
boundary surface. It should be remarked that $\rho=const$ is {\em not} an exact solution 
of the original non-expanded equation \eqref{eomrn}, so there must be some non-constant 
contributions to $\rho(x)$ at higher orders. However, up to the second order in $\epsilon$, 
only constant solution is permitted by the boundary conditions. 

Finally, we come to evaluate the area of the minimal surface using \eqref{mins}. Since now
$\rho'(x)=0$, we have
\be 
\mathcal{A}=2\pi \int_{x_0}^1 \mathrm{d}x\, g(\rho)^2 = 2\pi (1-x_0) r_B^2,
\ee
where $r_B=r_++\rho_0^2$ is the radius of the boundary surface.

\begin{figure}[tbp!]
\begin{center}
\includegraphics[width=.5\textwidth]{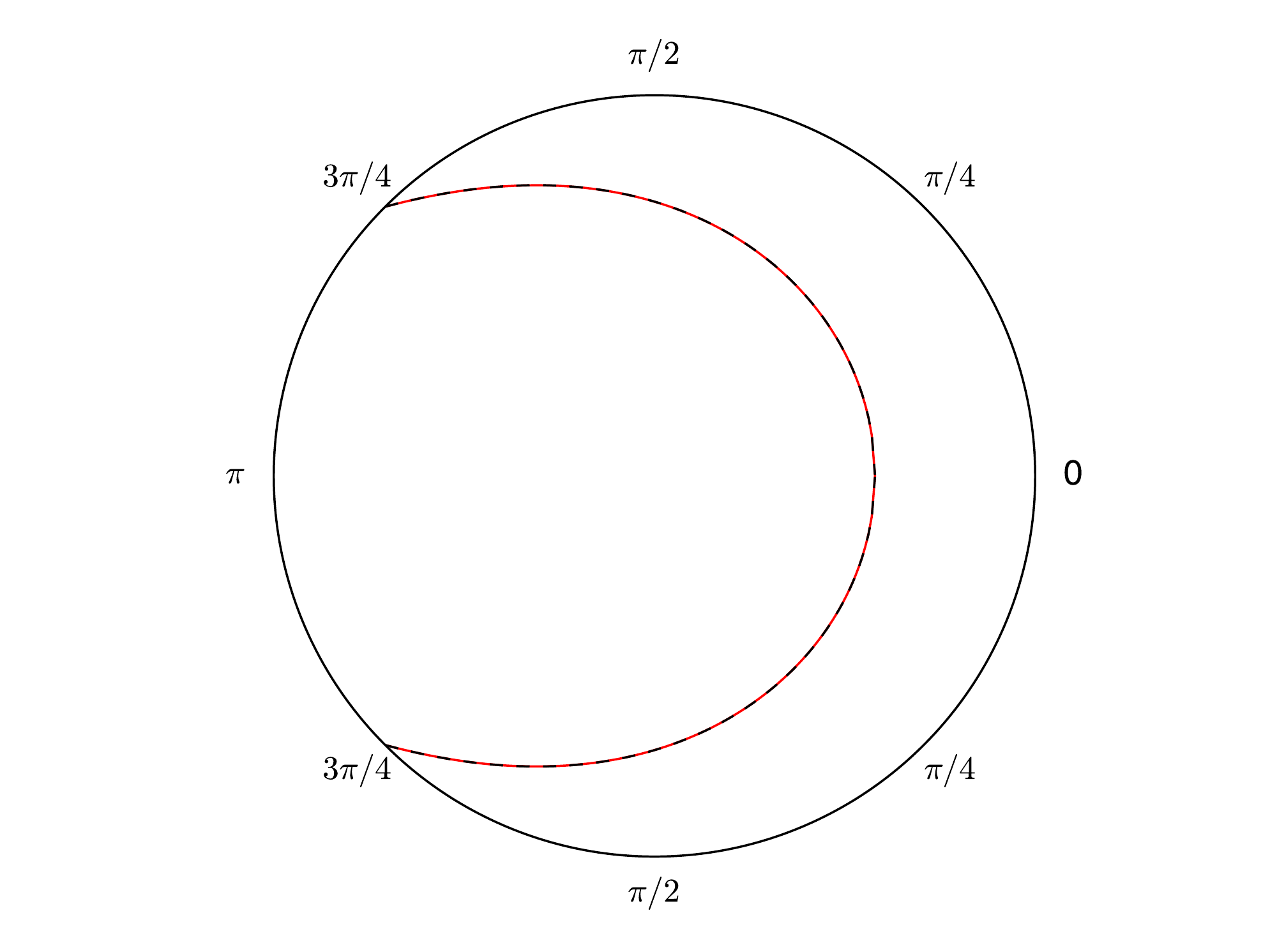}
\end{center}
\caption{The minimal surface in the case of a non-extremal RN black hole
with a near horizon boundary. The red curve is the numerical solution of (\ref{eomrn}). The 
dashed black curve is the perturbative solution (\ref{exp-rho}). The parameters are taken as: 
$\theta_0=3\pi/4,\rho_0=0.02,r_+=0.9,r_-=0.6.$} \label{fig2}
\end{figure}

\begin{figure}[htbp!]
\begin{center}
\includegraphics[width=.5\textwidth]{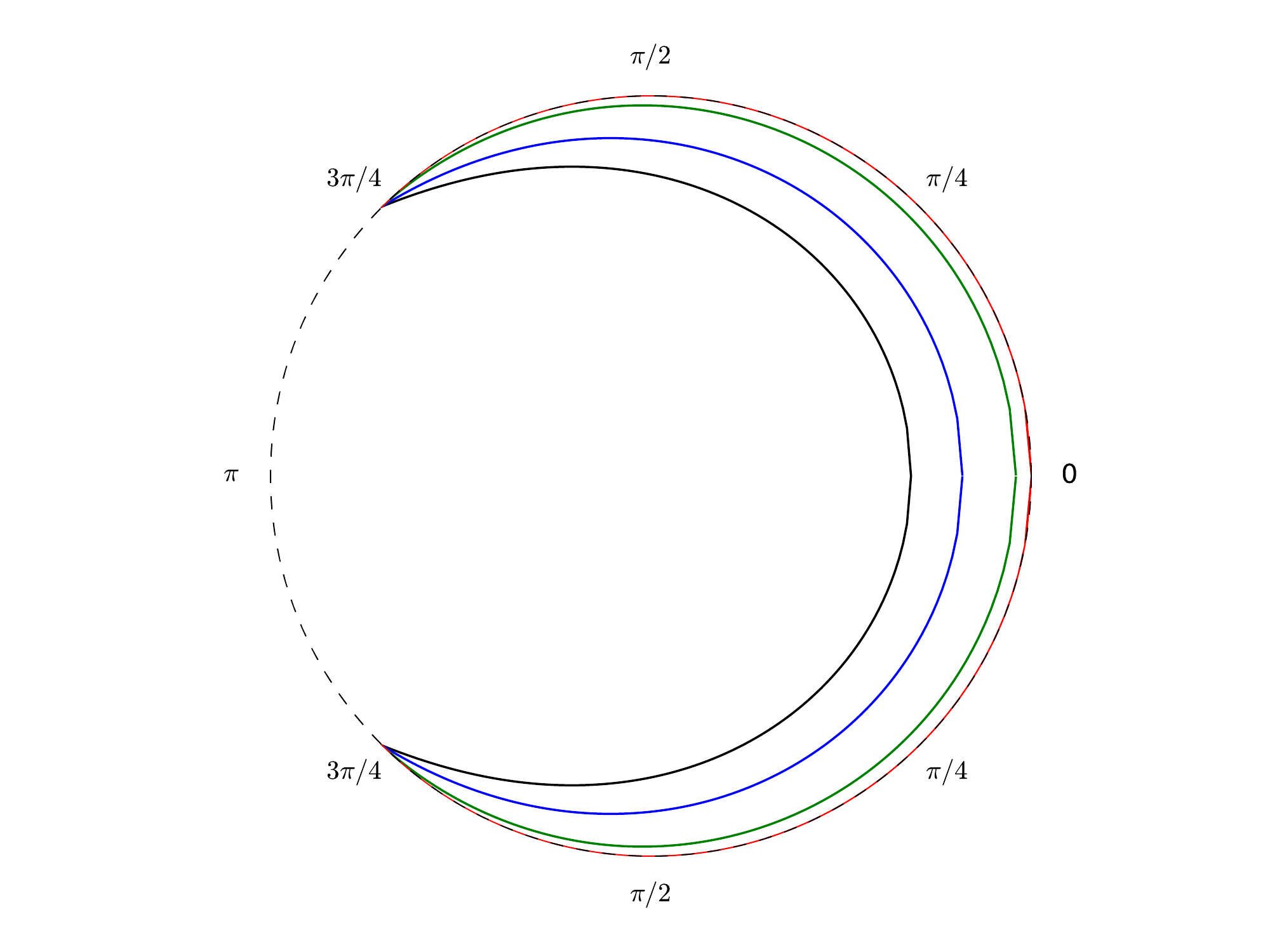}
\end{center}
\caption{The minimal surfaces of several RN black holes
with near horizon boundaries. Different curves correspond to different choices of 
the parameter $r_-$: $r_-=0.7$ (black), $r_-=0.8$ (blue), $r_-=0.88$(green), $r_-=0.9$(red).
The last case corresponds to extremal black hole, with numerical (red) and 
perturbative (dashed) results in good agreements.} \label{fig3}
\end{figure}

To verify the correctness of the perturbative procedures used in this section, we also 
studied the numerical solution to the equation \eqref{eomrn} for both the non-extremal
and extremal cases. The results are presented in Fig.\ref{fig2} and Fig.\ref{fig3}. 
These figures are created in the $(\rho,\theta)$ coordinates, with the $\varphi$ coordinate
omitted. The black hole 
event horizons are located at $\rho=0$ in such coordinates and are not demonstrated. 

Fig.\ref{fig2} corresponds to the case of a non-extremal RN black hole with a 
near horizon boundary (shown as the outer-most circle). The entangling surface is characterized 
by the single parameter $\theta_0=4\pi/3$. The other parameters are $\rho_0=0.02,
,r_+=0.9,r_-=0.6$.

Fig.\ref{fig3} gives the plot of minimal surfaces for several different RN black hole spacetimes.
The parameters $\theta_0, \rho_0$ and $r_+$ are identical to the case of Fig.\ref{fig2},
however several different values of $r_-$ are taken in order to indicate the change of 
minimal surface with respect to this parameter. Among these, the largest choice $r_-=0.9$
corresponds to the case of an extremal black hole. It can be seen that in both figures the 
perturbative results are in very good agreements with the numerical solutions.

\section{Conclusion}

In this paper, we studied the HEE associated with two asymptotically flat spacetimes, i.e.
Schwarzschild and RN spacetimes, with boundaries taken either at large but finite radial 
distances or close to the black hole event horizon. Here we summarize some of the interesting 
results:

\begin{itemize}
\item At the far end, the HEE for both spacetimes
are identical at the leading and the next to leading orders.
The leading order contribution is actually the HEE associated with the background Minkowski
spacetime, so the majority of the true contribution from the black holes comes in the next to 
leading order. It is shown that the next to leading order contribution to the HEE is always 
proportional to the black hole mass, regardless whether we are considering Schwarzschild or
RN spacetimes. This gives a version of the first law of HEE for asymptotically flat spacetimes.

\item Still at the far end, the case of RN spacetime begins to show off at the next to 
next to leading order $O(\epsilon^2)$. We have shown that there is an extra term which 
is proportional to $Q^2$ for the RN case. There are no conceivable differences between
the cases for extremal and non-extremal RN black hole spacetimes.

\item At the far end, we can compare the HEE of Schwarzschild black hole in flat spacetime with that of in AdS space. For the flat spacetime, the HEE up to the second order in $G M/r_\infty$ is 
\be 
S=\frac{r^2_\infty}{G}\left(a_0+a_1\frac{M}{r_\infty}+a_2\frac{M^2}{r_\infty^2}+...\right)
\ee
with $a_0=\pi\sin^2\theta_0,a_1=2\pi(1-\cos\theta_0)^2$, and $a_2$ can be read from equation (\ref{s2sch}), which dependent only on the size of subsystem $\theta_0$. For the case of Schwarzschild black hole in AdS$_4$, up to the first order in $M/L$ the HEE is (see Table 1 in \cite{1512.02816} where $L$ is omitted.)
\be 
S=\frac{L^2}{G}\left(b_0\frac{1}{\epsilon}-1+b_1\frac{M}{L}+...\right)
\ee
where $b_0=4\pi\sin\theta_0,b_1=8\pi(2\cos^3\theta_0-3\cos^2\theta_0+1)/3\sin2\theta_0$ depend only on the size of subsystem $\theta_0$, $\epsilon$ is a small cut-off of AdS boundary, and $L$ is curvature radius of AdS$_4$. Note that  in the above two expressions the background contributions to HEE, i.e. $M=0$, do not match with each other. However, we find that the sub-leading terms which contributed by the black hole(terms involving $M$) in two expressions are both scaled as $M/L$, if one choose $L\sim r_\infty.$ This choices is consistent with the fact that AdS space can be understood as a finite box with radius proportional to $L$ (Related discussion of this property in the context of gravitational collapse can be found in \cite{Cai:2016bsa}).

\item In the near horizon case, the HEE for Schwarzschild and non-extremal RN spacetimes 
are also quite similar: at the leading order, the HEE is proportional to the horizon area, 
and it actually equals to the horizon area if the subsystem is defined on the whole spherical 
boundary with a single point removed. Thus in this csae the leading term reproduces the black hole entropy. It is interesting to compare with results in \cite{Carlip:1998wz}, where the near horizon geometry is used to calculate the black hole entropy. The next to leading order contributions always vanish. 

\item Also in the near horizon case, the HEE for extremal and non-extremal RN spacetime 
behave quite differently. For the non-extremal case, the minimal surface anchored on the 
entangling surface on the boundary bends inwards, and hence the HEE is smaller than the area of 
the subsystem on the boundary divided by $4G$. However, for the extremal case, the minimal 
surface is completely immersed inside the boundary, up to the second order in the perturbative 
expansion. Therefore the HEE is equal  
to the area of the subsystem on the boundary divided by $4G$ on the same level of approximation.

\end{itemize}

Of course, the HEE associated with asymptotically flat spacetime is a subtle and relatively 
new subject and 
deserves further study. The content presented in this work should only be considered as 
initiating rather than closing this new area of study.

\section*{Acknowledgment}
This work is supported by the National Natural Science Foundation of China under the grant No. 11575088.


\begin{thebibliography}{100}
\bibitem{hep-th/9711200} 
  J.~M.~Maldacena,
  ``The Large N limit of superconformal field theories and supergravity,''
  Int.\ J.\ Theor.\ Phys.\  {\bf 38}, 1113 (1999)
  [Adv.\ Theor.\ Math.\ Phys.\  {\bf 2}, 231 (1998)]
  doi:10.1023/A:1026654312961
  [hep-th/9711200].


\bibitem{gr-qc/9310026} 
  G.~'t Hooft,
  ``Dimensional reduction in quantum gravity,''
  Salamfest 1993:0284-296
  [gr-qc/9310026].


\bibitem{hep-th/9409089} 
  L.~Susskind,
  ``The World as a hologram,''
  J.\ Math.\ Phys.\  {\bf 36}, 6377 (1995)
  doi:10.1063/1.531249
  [hep-th/9409089].


\bibitem{1010.3700} 
  W.~Li and T.~Takayanagi,
  ``Holography and Entanglement in Flat Spacetime,''
  Phys.\ Rev.\ Lett.\  {\bf 106}, 141301 (2011)
  doi:10.1103/PhysRevLett.106.141301
  [arXiv:1010.3700 [hep-th]].


\bibitem{hep-th/9906022} 
  R.~Bousso,
  ``Holography in general space-times,''
  JHEP {\bf 9906}, 028 (1999)
  doi:10.1088/1126-6708/1999/06/028
  [hep-th/9906022].


\bibitem{hep-th/0106113} 
  A.~Strominger,
  ``The dS / CFT correspondence,''
  JHEP {\bf 0110}, 034 (2001)
  doi:10.1088/1126-6708/2001/10/034
  [hep-th/0106113].


\bibitem{1603.05250} 
  F.~Sanches and S.~J.~Weinberg,
  ``Holographic entanglement entropy conjecture for general spacetimes,''
  Phys.\ Rev.\ D {\bf 94}, no. 8, 084034 (2016)
  doi:10.1103/PhysRevD.94.084034
  [arXiv:1603.05250 [hep-th]].


\bibitem{1611.02702} 
  Y.~Nomura, N.~Salzetta, F.~Sanches and S.~J.~Weinberg,
  ``Toward a Holographic Theory for General Spacetimes,''
  arXiv:1611.02702 [hep-th].


\bibitem{1410.4089} 
  A.~Bagchi, R.~Basu, D.~Grumiller and M.~Riegler,
  ``Entanglement entropy in Galilean conformal field theories and flat holography,''
  Phys.\ Rev.\ Lett.\  {\bf 114}, no. 11, 111602 (2015)
  doi:10.1103/PhysRevLett.114.111602
  [arXiv:1410.4089 [hep-th]].


\bibitem{1609.06203} 
  A.~Bagchi, R.~Basu, A.~Kakkar and A.~Mehra,
  ``Flat Holography: Aspects of the dual field theory,''
  arXiv:1609.06203 [hep-th].


\bibitem{hep-th/0603001} 
  S.~Ryu and T.~Takayanagi,
  ``Holographic derivation of entanglement entropy from AdS/CFT,''
  Phys.\ Rev.\ Lett.\  {\bf 96}, 181602 (2006)
  doi:10.1103/PhysRevLett.96.181602
  [hep-th/0603001].


\bibitem{hep-th/0605073} 
  S.~Ryu and T.~Takayanagi,
  ``Aspects of Holographic Entanglement Entropy,''
  JHEP {\bf 0608}, 045 (2006)
  doi:10.1088/1126-6708/2006/08/045
  [hep-th/0605073].


\bibitem{1304.4926} 
  A.~Lewkowycz and J.~Maldacena,
  ``Generalized gravitational entropy,''
  JHEP {\bf 1308}, 090 (2013)
  doi:10.1007/JHEP08(2013)090
  [arXiv:1304.4926 [hep-th]].


\bibitem{0705.0016} 
  V.~E.~Hubeny, M.~Rangamani and T.~Takayanagi,
  ``A Covariant holographic entanglement entropy proposal,''
  JHEP {\bf 0707}, 062 (2007)
  doi:10.1088/1126-6708/2007/07/062
  [arXiv:0705.0016 [hep-th]].


\bibitem{1310.5713} 
  X.~Dong,
  ``Holographic Entanglement Entropy for General Higher Derivative Gravity,''
  JHEP {\bf 1401}, 044 (2014)
  doi:10.1007/JHEP01(2014)044
  [arXiv:1310.5713 [hep-th]].


\bibitem{1306.4004} 
  V.~E.~Hubeny, H.~Maxfield, M.~Rangamani and E.~Tonni,
  ``Holographic entanglement plateaux,''
  JHEP {\bf 1308}, 092 (2013)
  doi:10.1007/JHEP08(2013)092
  [arXiv:1306.4004 [hep-th]].


\bibitem{1312.6887} 
  V.~E.~Hubeny and H.~Maxfield,
  ``Holographic probes of collapsing black holes,''
  JHEP {\bf 1403}, 097 (2014)
  doi:10.1007/JHEP03(2014)097
  [arXiv:1312.6887 [hep-th]].


\bibitem{1501.03019} 
  K.~Narayan,
  ``Extremal surfaces in de Sitter spacetime,''
  Phys.\ Rev.\ D {\bf 91}, no. 12, 126011 (2015)
  doi:10.1103/PhysRevD.91.126011
  [arXiv:1501.03019 [hep-th]].


\bibitem{1504.07430} 
  K.~Narayan,
  ``de Sitter space and extremal surfaces for spheres,''
  Phys.\ Lett.\ B {\bf 753}, 308 (2016)
  doi:10.1016/j.physletb.2015.12.019
  [arXiv:1504.07430 [hep-th]].


\bibitem{1512.02816} 
  N.~Kim and J.~Hun Lee,
  ``Time-evolution of the holographic entanglement entropy and metric perturbationst,''
  J.\ Korean Phys.\ Soc.\  {\bf 69}, no. 4, 623 (2016)
  doi:10.3938/jkps.69.623
  [arXiv:1512.02816 [hep-th]].

\bibitem{1212.1164} 
  J.~Bhattacharya, M.~Nozaki, T.~Takayanagi and T.~Ugajin,
  ``Thermodynamical Property of Entanglement Entropy for Excited States,''
  Phys.\ Rev.\ Lett.\  {\bf 110}, no. 9, 091602 (2013)
  doi:10.1103/PhysRevLett.110.091602
  [arXiv:1212.1164 [hep-th]].
  
\bibitem{1610.01542}
  D.~Momeni, M.~Faizal, S.~Bahamonde and R.~Myrzakulov,
  ``Holographic complexity for time-dependent backgrounds,''
  Phys.\ Lett.\ B {\bf 762}, 276 (2016)
  doi:10.1016/j.physletb.2016.09.036
  [arXiv:1610.01542 [hep-th]].

\bibitem{Carlip:1998wz} 
  S.~Carlip,
  ``Black hole entropy from conformal field theory in any dimension,''
  Phys.\ Rev.\ Lett.\  {\bf 82}, 2828 (1999)
  doi:10.1103/PhysRevLett.82.2828
  [hep-th/9812013].
\bibitem{Cai:2016bsa} 
  R.~G.~Cai, L.~W.~Ji and R.~Q.~Yang,
  ``On the critical behaviour of gapped gravitational collapse in confined spacetime,''
  arXiv:1609.02804 [gr-qc].
\end{thebibliography}

\providecommand{\href}[2]{#2}\begingroup
\footnotesize\itemsep=0pt
\providecommand{\eprint}[2][]{\href{http://arxiv.org/abs/#2}{arXiv:#2}}

\end{document}